\documentclass[12]{article}
\usepackage{graphicx}
\usepackage{lscape}
\usepackage{color}
\usepackage{url} 
\usepackage{amsmath}
\usepackage{amssymb}
\usepackage{natbib}
\usepackage{hyperref}

\begin{document}

\begin{center} {\bf Mean Squared Prediction Error Estimators of EBLUP of a Small Area Mean Under a Semi-Parametric Fay-Herriot Model} \end{center}

\date{}
\vspace{3mm}
\begin{center}
   B{\small Y} SHIJIE CHEN \\
   \vspace{2mm}
   {\it Vice President, Enanta Pharmaceuticals} \\
   \vspace{1mm}
   scip6153@gmail.com, \\
   \vspace{3mm}
   P. LAHIRI \\
   \vspace{2mm}
   {\it Professor, Joint Program in Survey Methodology \& Department of Mathematics, University of Maryland, College Park MD 20742, U.S.A. }\\
   \vspace{1mm}
   plahiri@umd.edu\\
   \vspace{3mm}
   A{\small ND} J.N.K. RAO \\
   \vspace{2mm}
   {\it Distinguished Research Professor, Carleton University, Ottawa, Ontario, Canada, K1S 5B6 }\\
   \vspace{1mm}
   jrao@math.carleton.ca
\end{center}

\vspace{3mm}

\begin{center}  {\large A}BSTRACT \end{center}

In this paper we derive a second-order unbiased (or nearly unbiased) mean squared prediction error
(MSPE) estimator of the empirical best linear unbiased predictor (EBLUP) of a small area mean for a
semi-parametric extension to the well-known Fay-Herriot model.  Specifically, we derive our MSPE
estimator essentially assuming certain moment conditions on both the sampling errors and random effects
distributions.  The normality-based Prasad-Rao MSPE estimator has a surprising robustness property
in that it remains second-order unbiased under the non-normality of random effects when a simple Prasad-Rao
method-of-moments estimator is used for the variance component and the sampling error distribution
is normal.  We show that the normality-based MSPE estimator is no longer second-order unbiased when
the sampling error distribution has non-zero kurtosis 
or when the Fay-Herriot moment method is used to
estimate the variance component, even when the sampling error distribution is normal.  It is
interesting to note that when the simple method-of moments estimator is used for the variance
component, our proposed MSPE estimator does not require the estimation of kurtosis of the random
effects. Results of a simulation study on the accuracy of the proposed MSPE estimator, under
non-normality of both sampling and random effects distributions, are also presented.

\vspace{5mm}

\noindent \emph{Some key words:} Mean Squared Prediction Errors;  Linear Mixed Model; Variance Components.

\addtolength{\baselineskip}{+1\baselineskip}

\vskip .2in

\begin{center}  {\large 1. I}NTRODUCTION \end{center}
 \hspace{4mm} In the context of estimating per-capita income for small places
(population less than 1000) from the 1970 Census of Population and Housing, \cite{FH1979} 
used a small area regression model and demonstrated that the resulting empirical best linear
unbiased predictors (EBLUP) have smaller average error than either the traditional survey estimators
or an alternative method using county averages. Let $Y_i$ be a direct estimator of the $i$th small
area mean $\theta_i$ and ${\bf x}_i=(x_{i1},\cdots,x_{ip})^{\prime}$ be a $p\times 1$ vector of
associated predictor variables, $i=1,\cdots,m.$  The Fay-Herriot model may be written as
$Y_i=\theta_i+e_i$ and $\theta_i={\bf x}_i^{\prime} {\boldsymbol\beta} + v_i, $ or as a linear mixed model:
$Y_i={\bf x}_i^{\prime} {\boldsymbol\beta}+v_i+e_i,$ where the sampling errors $\{e_i\}$ and the random effects
$\{v_i\}$ are independent with $e_i\stackrel{ind}\sim N(0,D_i) $ and $v_i\stackrel{ind}\sim
N(0,\psi),~i=1,\cdots,m.$ The sampling variances $D_i$ are assumed to be known, but ${\boldsymbol\beta}$ and
$\psi$ are to be estimated from the data $(Y_i,{\bf x}_i),~ i=1,\cdots,m.$ In practice, $D_i$'s are
externally estimated using the generalized variance function (GVF) method; see \cite{wol85}, \cite{FH1979}, \cite{otto1995sampling}, among others. The Fay-Herriot (FH) model has been used
extensively in  small area estimation and related problems for a variety of reasons, including its
simplicity, its ability to protect confidentiality of microdata and its ability to produce
design-consistent estimators [see \cite{rao2015small}, Chapter 7].

While the empirical best linear unbiased predictor (EBLUP) of a small area total or mean under the
FH model is easy to obtain, an accurate estimation of its mean square prediction error (MSPE) is a
challenging problem. A naive MSPE estimator is given by the MSPE of the best linear unbiased
predictor (BLUP) with the model variance $\psi$ replaced by a suitable estimator. But, it usually
underestimates the true MSPE of EBLUP mainly for two reasons. First, it fails to incorporate the
extra variability incurred due to the estimation of $\psi$ and the order of this underestimation is
$O(m^{-1}),$ for large $m$. Secondly, the naive MSPE estimator underestimates even the true MSPE of
the BLUP, the order of the underestimation being $O(m^{-1})$. \cite{prasad1990estimation} demonstrated
the importance of accounting for these two sources of underestimation, and using a Taylor
linearization method produced a second-order unbiased (or nearly unbiased) MSPE estimator of EBLUP
when the variance component is estimated by a simple method-of-moments. The bias of that MSPE
estimator is of order $o(m^{-1}).$ The derivation of their MSPE estimator involves essentially two
main steps. First a second-order correct MSPE expansion is obtained by neglecting all terms of
order $o(m^{-1})$. The second step involves the estimation of this second-order correct MSPE
approximation such that bias is of lower order, i.e. $o(m^{-1}).$ \cite{datta2000}  extended
the Prasad-Rao MSPE estimation method to maximum likelihood (ML) and restricted ML (REML) estimators of the model parameters. \cite{Datta2005} obtained a nearly unbiased MSPE estimator when $\psi$ is estimated by the \cite{FH1979} 
method-of-moments. \cite{DasJiangRao04}  generalized the Taylor method to general linear mixed
models and obtained nearly unbiased MSPE estimators.

Note that the derivation of EBLUP, using a moment estimator of $\psi$, does not require normality
assumptions. However, for the estimation of MSPE, \cite{prasad1990estimation}, \cite{datta2000},
\cite{DasJiangRao04},  \cite{Datta2005}, and others used the normality assumption.
One exception is the paper by \cite{lahiri1995robust}  who assumed normality of the sampling errors
$\{ e_i\},$ but replaced the normality of the random effects $\{ v_i\}$ by certain moment
conditions. They showed that the normality-based Prasad-Rao estimator of the MSPE of EBLUP remains
second-order unbiased under this non-normal set-up when the simple method-of-moments estimator of
$\psi$ is used. This is indeed a surprising result, demonstrating the robustness of the Prasad-Rao
MSPE estimator under unspecified non-normality of the random effects $\{ v_i\}$. Does this result
hold when the sampling errors $\{ e_i\}$ are non-normal or when $\psi$ is estimated by the moment
estimator of $\psi$ proposed by \cite{FH1979}?
The case of non-normal sampling errors may be useful when the sample sizes are small for several areas.  In the latter case, the effect of the central limit theorem may not hold.

In Section 2, we introduce 
a semi-parametric Fay-Herriot model that assumes certain moment conditions and briefly review the empirical best linear unbiased
prediction (EBLUP).  In Section 3, we obtain a second-order approximation
to the MSPE of the EBLUP without the normality assumption.  In Section 4,
we propose a second-order (or nearly) unbiased MSPE estimator of
EBLUP, using the approximation to the MSPE obtained in Section 3.
Finally, some empirical results from a small simulation study are
reported in Section 5. Proofs of our results are sketched in the
Appendix.

\begin{center}  {\large 2. A SEMI-PARAMETRIC FAY-HERRIOT MODEL AND EBLUP } \end{center}

We introduce a semi-parametric Fay-Herriot model
as $Y_i=\theta_i+e_i$ and
$\theta_i={\bf x}_i^{\prime}\boldsymbol\beta+v_i, ~i=1,\cdots,m,$ or as a linear mixed model,
\begin{eqnarray}\label{fh}
Y_i&=&{\bf x}_i^{\prime}\boldsymbol{\beta}+v_i+e_i\;,\;\;\;i=1,\cdots,m\;,
\end{eqnarray}
where the sampling errors $\{e_i\}$ and the random effects $\{v_i\}$ are uncorrelated with $e_i
{\sim}[0, D_i, \kappa_{ei}]$ and  $v_i {\sim}[0, \psi,\kappa_v ],$ $[\mu,\sigma^2,\kappa]$
representing a probability distribution with mean $\mu$, variance $\sigma^2$ and kurtosis $\kappa.$
We define kurtosis of a distribution as $\kappa=\mu_4/\sigma^4-3,$ where $\sigma^2$ and $\mu_4$ are
the variance and the fourth central moment of the distribution, respectively.  We assume that
$\boldsymbol{\beta}, \psi,$ and $\kappa_v$ are unknown. In addition to the conventional assumption of the normality-based Fay-Herriot model for the sampling variance, which posits that the sampling variance 
$D_i$ is known, we also assume that the sampling kurtosis $\kappa_{ei}$ is known.


Define ${\bf X}^{\prime}=({\bf x}_1,\cdots, {\bf x}_m),$ and ${\bf \Sigma}(\psi)=diag\{\psi+D_j; j=1,...,m\}$ and $\boldsymbol\beta$
can be estimated by $\hat{\boldsymbol\beta}(\psi)=
[{\bf X}^{\prime}{\bf \Sigma}^{-1}(\psi){\bf X}]^{-1}{\bf X}^{\prime}{\bf \Sigma}^{-1}(\psi){\bf Y},$ a weighted least square estimator
of $\boldsymbol\beta$ for a given $\psi.$ The BLUP of $\theta_i$ under the FH model (\ref{fh}) is given by:
$$\hat \theta_i(Y_i; \psi)= B_iY_i+(1-B_i){\bf x}^{\prime}_i\hat
{\boldsymbol\beta}(\psi),$$ where  $B_i=\frac{\psi}{D_i+\psi},\;i=1,\ldots,m.$  An EBLUP of $\theta_i$ is then
obtained as
$$\hat \theta_i(Y_i; \hat \psi)= \hat B_iY_i+(1-\hat
B_i){\bf x}^{\prime}_i\hat {\boldsymbol\beta}(\hat \psi)=:\hat{\theta}_i,$$ where $\hat B_i=\frac{\hat \psi}{D_i+\hat
\psi}, ~i=1,\ldots,m,$ and $\hat{\psi}$ is a moment estimator of $\psi.$ Note that the BLUP and
EBLUP, based on $\hat\psi,$ do not require normality of $\{e_i\}$ and $\{v_i\}$.

\cite{prasad1990estimation} proposed the following simple method-of-moments estimator of $\psi$:
\begin{eqnarray*}
\hat{\psi}_{PR}= max\left \{0,(m-p)^{-1} \sum_{j=1}^m \big \{ (Y_j-{\bf x}_j^{\prime}\hat{\boldsymbol\beta}_{\rm
OLS})^2 -(1-h_{jj})D_j\big \}\right \},
\end{eqnarray*}
 where $\hat{\boldsymbol\beta}_{\rm OLS} =\big
({\bf X}^{\prime}{\bf X} \big )^{-1} {\bf X}^{\prime}{\bf Y},$ the ordinary least square estimator of ${\boldsymbol\beta},$ and $
h_{jj}={\bf x}_j^{\prime} \big ({\bf X}^{\prime}{\bf X}\big )^{-1}{\bf x}_j,$ the leverage for the $j$th small area
$(j=1,\cdots,m).$  \cite{FH1979} obtained a different moment estimator $\hat{\psi}_{FH},$
by solving the following equation iteratively for $\psi$:
\begin{eqnarray}
A(\psi)=\frac{1}{m-p}{\bf Y}'{\bf Q}(\psi){\bf Y}-1=0,
\end{eqnarray}
where
$${\bf Q}(\psi)={\bf \Sigma}^{-1}(\psi)-{\bf \Sigma}^{-1}(\psi){\bf X}\{{\bf X}'{\bf \Sigma}^{-1}(\psi){\bf X}\}^{-1}{\bf X}'{\bf \Sigma}^{-1}(\psi),$$
and ${\bf Y}'{\bf Q}(\psi){\bf Y}=\sum_{j=1}^m(D_j+\psi)^{-1}\{Y_j-{\bf x}_j^{\prime}\hat
{\boldsymbol\beta}(\psi)\}^2$ 
is the weighted residual sum of squares.
\cite{pn81} proposed a similar moment method in
the context of regression analysis of survey data.  For the
special case $D_i=D\; (i=1,\cdots,m)$,
$\hat{\psi}_{PR}=\hat{\psi}_{FH}.$  In this paper, we focus on
EBLUP based on $\hat{\psi}_{PR}$ or $\hat{\psi}_{FH}.$
The estimators $\hat{\psi}_{PR}$  and $\hat{\psi}_{FH}$  are typically consistent estimators for
large $m$ under the following regularity conditions:
\begin{description}
\item $(r.1) ~0 < D_L \le D_j \le D_U < \infty, ~j=1,\cdots,m,$
\item $(r.2)~ \sup_{j\ge 1} h_{jj} = O(\frac{1}{m});$
\end{description}
i.e., $\hat\psi-\psi=O_p(m^{-\frac{1}{2}})$ for
$\hat\psi=\hat\psi_{PR}$ and $\hat\psi_{FH}$.

Under non-normality and regularity conditions, the bias of $\hat \psi_{PR}$ is of order
$o(m^{-1}).$ However, unless $D_i=D\; (i=1,\cdots,m)$, the bias of $\hat \psi_{FH}$ is of order
$O(m^{-1}),$ even under normality, and it is given by
\begin{eqnarray}
E[\hat \psi_{FH} -\psi]=b(\hat\psi_{FH};\psi,\kappa_v) + o(m^{-1}),
\end{eqnarray}
where
\begin{eqnarray*}
b(\hat \psi_{FH};\psi,\kappa_v)&=&b_N(\hat \psi_{FH};\psi) +\alpha(\hat \psi_{FH};\psi,\kappa_v
)\\
b_N(\hat \psi_{FH};\psi)&=&\frac { 2[m {\rm tr} ({\bf \Sigma}^{-2}) -\{ {\rm tr}({\bf \Sigma}^{-1})\}^2]}
{[{\rm tr} ({\bf \Sigma}^{-1})]^3}\\
\alpha(\hat \psi_{FH};\psi,\kappa_v )
                      &=&\frac {[{\rm tr}({\bf \Sigma}^{-2})]^2- {\rm tr}({\bf \Sigma}^{-3}){\rm tr}({\bf \Sigma}^{-1})}
                                     {[{\rm tr}({\bf \Sigma}^{-1})]^3}\psi^2\kappa_v
                         +\frac{  {\rm tr}({\bf D}^2{\bf \Phi}{\bf \Sigma}^{-2}){\rm tr}({\bf \Sigma}^{-2})-{\rm tr}({\bf \Sigma}^{-1})
                                      {\rm tr}({\bf D}^2{\bf \Phi}{\bf \Sigma}^{-3})}
                                {[{\rm tr}({\bf \Sigma}^{-1})]^3},\\
                                {\bf D}&=&diag\{D_j; j=1,\cdots,m\},\\
                                {\bf \Phi}&=&diag\{\kappa_{ej}; j=1,\cdots,m\}.
\end{eqnarray*}
The proof of (3) is given in the Appendix. In the above, $b_N(\hat \psi_{FH}; \psi)$ denotes  the bias of $\hat \psi_{FH}$ up to order
$O(m^{-1})$ under normality [see \cite{Datta2005}] and $\alpha( \hat\psi_{FH};
\psi,\kappa_v)$ is the additional non-normality effect on the bias.

The variance of $\hat\psi$ may be expressed as
\begin{eqnarray}\label{varpsi}
var(\hat \psi)=var_N(\hat \psi)+\eta( \hat\psi; \psi,\kappa_v),
\end{eqnarray}
where $var_N(\hat \psi)$ denotes the variance of $\hat \psi$ up to order $O(m^{-1})$ under
normality, and $\eta({\hat \psi}, \psi,\kappa_v)$ denotes the additional non-normality effect on
the variance. From \cite{Datta2005}, we have
\begin{eqnarray*}
var_N(\hat \psi_{PR})&=&2m^{-2}\sum_{j=1}^m(\psi+D_j)^2=2m^{-2}{\rm tr}({\bf \Sigma}^{{2}}),\\
var_N(\hat \psi_{FH})&=&2m\left \{\sum_{j=1}^m(\psi+D_j)^{-1}\right \}^{-2}=2m \left \{ {\rm tr}(
{\bf \Sigma}^{-1}) \right \}^{-2}.
\end{eqnarray*}
It can be shown that
\begin{eqnarray*}
\eta( \hat \psi_{PR};  \psi,\kappa_v)&=& m^{-1} \left
\{\kappa_v\psi^2+\frac{1}{m}\sum_{j=1}^m\kappa_{ej}D_j^{2}\right \}
=m^{-1} \left \{\kappa_v\psi^2+ m^{-1} {\rm tr}({\bf D}^2{\bf \Phi}) \right \} \\
\eta( \hat \psi_{FH};  \psi,\kappa_v)&=&\left \{\sum_j(\psi+D_j)^{-1}\right \}^{-2}
 \sum_{j=1}^m\left\{(\psi+D_j)^{-2}[\kappa_v\psi^2+\kappa_{ej}D_j^{2}]\right\}\\
 &= &[{\rm tr}({\bf \Sigma}^{-1})]^{-2}\left\{ {\rm tr}({\bf \Sigma}^{-2}) \kappa_v\psi^2+  {\rm
 tr}({\bf D}^2{\bf \Phi}{\bf \Sigma}^{-2})\right\}.
\end{eqnarray*}
See Appendix for a proof of the asymptotic variance of $\hat \psi$.  When both $\{ e_i\}$ and $\{ v_i\}$ are normal, we have $var(\hat \psi_{PR})=var_N(\hat
\psi_{PR}),\;var(\hat \psi_{FH})=var_N(\hat \psi_{FH})$ and  $var_N(\hat \psi_{FH}) \le var_N(\hat
\psi_{PR})$  with equality for the balanced case $D_i=D,~(i=1,\cdots,m),$  see  \cite{Datta2005}.  It is interesting to note that the latter result does not extend to the 
situation with non-zero kurtosis. For the balanced case $D_i=D~(i=1,\cdots,m),$ we have $b(\hat
\psi_{PR};\psi,\kappa_v)=b(\hat \psi_{FH};\psi,\kappa_v)=0$ and $var(\hat \psi_{PR})=var(\hat
\psi_{FH})=2m^{-1}(\psi+D)^2+m^{-1}\{\kappa_v\psi^2+m^{-1}D^2{\rm tr}({\bf \Phi})\}$ simply because in
this situation $\hat{\psi}_{PR}=\hat{\psi}_{FH}.$

\begin{center}  {\large 3. A}PPROXIMATION TO MSPE \end{center}

The MSPE of the EBLUP $\hat{\theta}_i$ is given by $MSPE(\hat {\theta}_i)=E(\hat
{\theta}_i-\theta_i)^2,$ where the expectation is taken over the marginal distribution of ${\bf Y}$ under
the semi-parametric Fay-Herriot model (\ref{fh}). The MSPE of the BLUP of $\hat\theta_i ( Y_i,\psi)$ is not
affected by non-normality and it is given by
$$MSPE[\hat\theta_i ( Y_i,\psi)]= g_{1i}(\psi)+g_{2i}(\psi),$$
where
\begin{eqnarray*}
g_{1i}(\psi)&=&\frac{\psi D_i}{\psi+D_i},\\
g_{2i}(\psi)&=&\frac{D_i^{2}}{(\psi+D_i)^2}
var[{\bf x}'_i\hat{\boldsymbol\beta}(\psi)]=\frac{D_i^{2}}{(\psi+D_i)^2}{\bf x}_i^{\prime}\left
[{\bf X}^{\prime}{\bf \Sigma}^{-1}(\psi){\bf X}\right ]^{-1}{\bf x}_i.
\end{eqnarray*}
A naive estimator of $MSPE[\hat\theta_i ( Y_i,\psi)]$ that does not incorporate variability incurred due to estimation of $\psi$ is given by
$$mspe_i^{naive}=g_{1i}(\hat\psi)+ g_{2i}(\hat\psi).$$

We are interested in approximating the MSPE of the EBLUP under the proposed semi-parametric model (1)
that accounts for the estimation of $\psi$ and is
second-order accurate, i.e., accurate up to order $O(m^{-1})$.

We decompose the MSPE of the EBLUP $\hat{\theta}_i$ as
\begin{eqnarray}\label{decompose}
MSPE[\hat\theta_i ( Y_i,\hat\psi)]&=&MSPE[\hat\theta_i ( Y_i,\psi)]+ E[ \hat\theta_i (
Y_i,\hat\psi)- \hat\theta_i ( Y_i,\psi) ]^2\nonumber\\
&&+2E[\hat\theta_i ( Y_i,\hat \psi)- \hat\theta_i ( Y_i,\psi) ][\hat\theta_i ( Y_i,\psi)-\theta_i].
\end{eqnarray}
where $\hat\theta_i ( Y_i,\hat\psi)=\hat{\theta}_i$ and $\hat\theta_i ( Y_i,\psi)$ is the BLUP. The
cross-product term in (\ref{decompose}) is zero under normality of $\{v_i\}$ and $\{e_i\};$ see \cite{kac84}, but it is of $O(m^{-1})$ under the proposed semi-parametric FH model (1) and hence not negligible.
Result 1, given below, provides the following approximations to the last two terms of (\ref{decompose}).  A proof of Result 1 is given in the Appendix.

\noindent{ \bf Result 1:} Under the proposed semi-parametric  FH model (\ref{fh}) and
regularity conditions (r.1), (r.2) and\\
(r.3): $\sup_{j \ge 1} E\mid v_j\mid^{8+\delta}<\infty,\; {\rm
0<\delta<1}$, we have
\begin{eqnarray*}
(i)& &E[ \hat\theta_i ( Y_i,\hat\psi) -\hat\theta_i ( Y_i,\psi)  ]^2=g_{3i}(\psi,\kappa_v)+o(m^{-1}),\\
(ii)& &E[\hat\theta_i ( Y_i,\hat\psi)- \hat\theta_i ( Y_i,\psi)
][\hat\theta_i ( Y_i,\psi)
-\theta_i]=g_{4i}(\psi,\kappa_v)+o(m^{-1}),
\end{eqnarray*}
where
\begin{eqnarray*}
 g_{3i}(\psi,\kappa_v)&=&\frac{D_i^{2}}{(\psi+D_i)^3}var(\hat \psi),\\
 g_{4i}(\psi,\kappa_v)&=&\frac{\psi D_i^{2}}{m(\psi+D_i)^3}\left[D_i\kappa_{ei}-\psi\kappa_v\right
 ]c(\hat \psi;\psi),
\end{eqnarray*}
$c(\hat \psi_{PR};\psi)=1$ and $c(\hat \psi_{FH};\psi)=m(\psi+D_i)^{-1}\left \{\sum_j
(\psi+D_j)^{-1}\right \}^{-1}.$

Thus, a second-order expansion to MSPE of the EBLUP $\hat{\theta}_i$ is given by
\begin{eqnarray}\label{amspe}
& &AMSPE_i\nonumber\\
&=&g_{1i}(\psi)+g_{2i}(\psi)+g_{3i}(\psi,\kappa_v)+2g_{4i}(\psi,\kappa_v)\nonumber\\
&=&\frac{\psi D_i}{\psi+D_i}+\frac{D_i^{2}}{(\psi+D_i)^2} var[{\bf x}'_i\hat{\boldsymbol\beta}(\psi)]+
                              \frac{D_i^{2}}{(\psi+D_i)^3}var(\hat \psi)+
\frac{2\psi D_i^{2}}{m(\psi+D_i)^3}\left[D_i\kappa_{ei}-\psi\kappa_v\right ]c(\hat \psi;\psi)\nonumber\\
                              &=&AMSPE_{i,N}+\frac{D_i^{2}}{(\psi+D_i)^3}\eta( \hat\psi; \psi,\kappa_v)+2g_{4i}(\psi,\kappa_v),                
\end{eqnarray}
where $AMSPE_{i,N}$ is the normality-based MSPE approximation as given in \cite{prasad1990estimation} and
\cite{Datta2005}. The term $g_{3i}(\psi,\kappa_v)$ is the additional uncertainty due to
the estimation of the variance component $\psi$ and the term $2g_{4i}(\psi,\kappa_v)$ is needed to
adjust for non-zero kurtosis. Under the regularity conditions, $ g_{1i}(\psi)$ is the leading term
[of order $O(1)$] and the remaining terms are all of order $O(m^{-1}).$ Note that non-normality
affects both $var(\hat \psi)$ and the cross-product term $2E[\hat\theta_i(\hat
\psi,{\bf Y})-\hat\theta_i(\psi,{\bf Y})][\hat\theta_i(\psi,{\bf Y})-\theta_i]$. When both $\{ e_i\}$ and $\{ v_i\}$
are normal, the above approximation reduces to the Prasad-Rao (1990) approximation when
$\hat\psi=\hat\psi_{PR}$ and the \cite{Datta2005} approximation when
$\hat\psi=\hat\psi_{FH}$.   When the $\{ e_i\}$ are normal and $\hat\psi=\hat\psi_{PR}$, the MSPE
approximation (\ref{amspe}) reduces to the Lahiri-Rao (1995) approximation.

\begin{center}  {\large 4. N}EARLY UNBIASED ESTIMATOR OF MSPE \end{center}

The second-order MSPE approximation $AMSPE_i,$ given by (\ref{amspe}),
involves unknown parameters $\psi$ and $\kappa_v$. Let
$\hat\kappa_v$ be a consistent estimator of $\kappa_v$.  Then
$g_{2i}(\hat \psi),~ g_{3i}(\hat \psi,\hat\kappa_v),$ and  $
g_{4i}(\hat \psi,\hat\kappa_v)$ are second-order unbiased (or nearly
unbiased) estimators of $g_{2i}( \psi),~ g_{3i}( \psi,\kappa_v),$
and  $g_{4i}( \psi,\kappa_v)$, respectively, since latter functions of
$\psi$ and $\kappa_v$ are already of order $O(m^{-1})$. However,
estimation of the the leading term $g_{1i}( \psi)$ in (\ref{amspe}) needs
special attention since it is of the order $O(1).$

Under regularity conditions (r.1) and (r.2), it can be shown that
\begin{eqnarray}\label{g1}
E[g_{1i}(\hat \psi)]&=&g_{1i}( \psi)-g_{3i}(\psi,\kappa_v)+g_{5i}(\psi,\kappa_v)+o(m^{-1}),
\end{eqnarray}
where $g_{5i}(\psi,\kappa_v)=\frac{D_i^{2}}{(\psi+D_i)^2}b( \hat \psi; \psi,\kappa_v).$ Using (\ref{amspe})
and (\ref{g1}), a second-order unbiased (or nearly unbiased) MSPE estimator is then given by
\begin{eqnarray}\label{mspi}
mspe_i= g_{1i}(\hat \psi)+g_{2i}(\hat \psi)+2g_{3i}(\hat \psi,\hat\kappa_v)+2g_{4i}(\hat
\psi,\hat\kappa_v )- g_{5i}(\hat\psi,\hat\kappa_v).
\end{eqnarray}

When $\hat \psi=\hat \psi_{PR},$ we have
\begin{eqnarray}\label{mspepr}
mspe_i^{PR}&= &mspe_{i,N}^{PR}+\frac{2D_i^{2}}{m(\hat \psi+D_i)^3} \left [\hat \psi D_i\kappa_{ei}+
\frac{1}{m}\sum_{j=1}^m\kappa_{ej}D_j^{2}\right ]\nonumber\\
&=&mspe_{i,N}^{PR}+\frac{2D_i^{2}}{m(\hat \psi+D_i)^3} \left [\hat \psi D_i\kappa_{ei}+ m^{-1} {\rm
tr} ({\bf D}^2{\bf \Phi}) \right ],
\end{eqnarray}
where $mspe_{i,N}^{PR}=g_{1i}(\hat \psi)+g_{2i}(\hat \psi)+\frac{2D_i^2}{m^2(\hat
\psi+D_i)^3}\sum_{j=1}^m(\hat \psi+D_j)^2$ is the normality-based MSPE estimator first proposed by
\cite{prasad1990estimation}.   It is interesting to note that the MSPE estimator (\ref{mspepr}) does not require the
estimation of $\kappa_v$ although $\kappa_v$ is involved in the MSPE approximation (\ref{amspe}). For normal
$\{ e_i\}$ but unspecified non-normal $\{ v_i\}$, $mspe_i^{PR}=mspe_{i,N}^{PR}$  (Lahiri and Rao,
1995). It is interesting to note that the robustness of the Prasad-Rao MSPE estimator does not
extend to the case  when $\{ e_i\}$ have non-zero kurtosis.  When $\kappa_{ej}>0,~j=1,\cdots, m$,
$mspe_i^{PR}$ overestimates whenever $mspe_{i,N}^{PR}$ overestimates.

When $\hat \psi=\hat \psi_{FH}$, we have
\begin{eqnarray}\label{mspefh}
mspe_i^{FH}=mspe_i^{DRS}+\frac{2D_i^{2}}{(\hat \psi_{FH}   +D_i)^3}\eta (\hat \psi_{FH};\hat
\psi_{FH},\hat\kappa_v)+2g_{4i}(\hat \psi_{FH},\hat
\kappa_v)-\frac{D_i^{2}}{(\hat\psi_{FH}+D_i)^2}\alpha(\hat \psi_{FH}; \hat \psi_{FH},\hat\kappa_v)\nonumber\\
\end{eqnarray}
where
\begin{eqnarray*}
mspe_i^{DRS}=g_{1i}(\hat \psi_{FH})+g_{2i}(\hat \psi_{FH})+\frac{2D_i^2}{(\hat
\psi_{FH}+D_i)^3}evar_N(\hat \psi_{FH})-\frac{D_i^{2}}{(\hat \psi_{FH}+D_i)^2} {b}_N (\hat
\psi_{FH}; \hat \psi_{FH})
 \end{eqnarray*}
is the normality-based estimator proposed by \cite{Datta2005}.  In the above
$evar_N(\hat \psi_{FH})$ is obtained from $var_N(\hat \psi_{FH})$ using $\hat \psi_{FH}$ in place
of $\psi$. When both $\{e_i\}$ and $\{ v_i\}$ are normal, $mspe_i^{FH},$ given by (\ref{mspefh}), reduces to
$mspe_i^{DRS}.$ However, when $\{ e_i \}$ are normal but not $\{ v_i \}$, $mspe_i^{FH}$ is not
identical to $mspe_i^{DRS}$ unless $D_i=D\;(i=1,\cdots,m).$ Hence $mspe_i^{DRS},$ unlike
$mspe_i^{PR},$ is not robust under non-normality of $\{ v_i\}$ , even when
 $\{ e_i \}$ are normal.   It is easy to check that for the balanced case $D_i=D\;(i=1,\cdots,m),$
$mspe_i^{FH}=mspe_i^{PR}.$

%

To obtain a consistent estimator of $\kappa_v$, we replace $\psi$
and $var(\hat\psi_{FH})$ in (\ref{varpsi}) by $\hat\psi_{FH}$ and
$$v_{WJ}=\sum_{u=1}^mw_u(\hat\psi_{FH,(-u)}-\hat\psi_{FH})^2,$$ 
a weighted jackknife estimator of $var(\hat\psi_{FH})$ considered by
\cite{cl08},
 where  $w_u=1-h_{uu}$ and $\hat\psi_{FH,(-u)}$ is the
Fay-Herriot estimator of $\psi$, using all but the $u$th small area data.  The resulting equation
is given by:
\begin{eqnarray}\label{modelkur}
k(\hat\psi_{FH})+l(\hat\psi_{FH})\kappa_v=0,
\end{eqnarray}
where
\begin{eqnarray*}
k(\psi)&=&2m+{\rm tr}({\bf D}^2{\bf \Phi}{\bf \Sigma}^{-2})-\{{\rm tr}({\bf \Sigma}^{-1})\}^2v_{WJ}\\
l(\psi)&=&{\rm tr}({\bf \Sigma}^{-2})\psi^2.
\end{eqnarray*}
Solving (\ref{modelkur}) for $\kappa_v$, we obtain a closed-form estimator of
$\kappa_v:$ $\hat \kappa_v=-k( \hat \psi_{FH} )/l(\hat \psi_{FH})$
if $\hat \psi_{FH}>0$ and $0$ otherwise. Since this is a smooth
function and $\hat \psi_{FH}$ and $v_{WJ}$ are consistent for
$\psi$ and $var(\hat\psi_{FH})$ respectively, we have consistency
of the estimator $\hat\kappa_v$.

\begin{center}  {\large 5. S}IMULATION STUDY \end{center}

Finite-sample accuracy of the proposed robust estimator of the MSPE  of the EBLUP is investigated in this section through a Monte Carlo
simulation study, for the special case ${\bf x}_i^{\prime}\boldsymbol\beta=\mu$ and
$\kappa_{ei}=\kappa_e ~(i=1,\cdots,m).$  Noting that the MSPE is
translation invariant (i.e., it remains the same when $Y_i$ is
changed to $Y_i-\mu$), we set $\mu=0$ without loss of generality.
However, to account for the uncertainty in the estimation of the
common mean that arises in practice, we still estimate the zero
mean.  We have considered different values of $m$, the number of
small areas.  But, to save space, we report the simulation results
only for $m=60$ and summarize results for other values of $m$.
Further, we consider nine combinations of $\kappa_v=0,3,6$ and
$\kappa_e=0,3,6.$ Note that in our simulation $\kappa=0,3,$ and
$6$ correspond to the normal, double exponential, and {\it shifted}
exponential with mean zero distributions, respectively.  We set $\psi=1.$
Regarding the known sampling variances, we consider both a
balanced case, i.e. $D_j=D=\psi\; (j=1,\cdots,m)$ and an
unbalanced case that corresponds to the Type II pattern of \cite{Datta2005}. Thus, for the unbalanced case, there are 5
groups $G_t,\;(t=1,\cdots,5)$ of equal number of small areas such that within each
group sampling variances are the same. Specifically,
$D_j=2.0\;j\in G_1;\;D_j=0.6\;j\in G_2;\;D_j=0.5\;j\in
G_3;\;D_j=0.4\;j\in G_4;\;D_j=0.2\;j\in G_5.$

For both the balanced and unbalanced cases, we compared the
performances of our proposed robust MSPE estimator of the EBLUP based
on the Prasad-Rao simple method-of-moments estimator of $\psi$
with the naive and the normality-based Prasad-Rao (PR) MSPE
estimator. In addition, for the unbalanced case, we compared the
performance of our proposed robust MSPE estimator of the EBLUP based
on the Fay-Herriot method-of-moments estimator with the naive and the
normality-based Datta-Rao-Smith (DRS)  MSPE estimators.  Note that
in the balanced case $\hat{\psi}_{PR}=\hat{\psi}_{FH}$ and in this
case our robust method does not involve estimation of the unknown
kurtosis $\kappa_v.$ For the unbalanced case, estimation of
$\kappa_v$ is not needed for the Prasad-Rao method of estimating
$\psi$, but it is needed for the Fay-Herriot method of estimating
$\psi$.

For each case with specified parameters, we generated $R=10,000$
independent set of variates $\{ v_i,e_i,~ i=1,\cdots,m\}$. Simulated values of the relative bias (RB) and of the relative root mean
squared error (RRMSE) of estimators of mean squared prediction error of the 
EBLUP were then
computed for all the $m$ areas  as follows:
\begin{eqnarray*}
RB_i&=&100\frac{E(mspe_i)-MSPE_i}{MSPE_i},\\
RRMSE_i&=&100\frac{\sqrt{E(mspe_i-MSPE_i)^2}}{MSPE_i}
\end{eqnarray*}
where $mspe_i$ denotes an estimator of $MSPE_i,$ the MSPE of the EBLUP
$\hat {\theta}_i(y_i;\hat A)$ of the true small-area mean
$\theta_i~(i=1,\cdots,m),$ and the expectation ``{\it E}" is
approximated by the Monte Carlo method.  The RB and RRMSE, averaged over areas with the same sampling variances, were then
reported in Tables 1-4.

%

Table 1 reports the percent RB of MSPE estimators for the balanced
case. For each MSPE estimator, the absolute RB decreases as the
number of small areas, $m$, increases. For all the cases, the
naive estimator leads to underestimation, ranging from about $7\%$
to $23\%$ for $m=30$ and about $3\%$ to $14\%$ for $m=60.$ Results
for the Prasad-Rao MSPE estimator and the proposed MSPE estimator
are almost identical and RB is negligible whenever the sampling
errors $\{e_i\}$ are normally distributed; this is consistent with
 previous theory (\cite{lahiri1995robust}).  When the $\{e_i\}$
are non-normal, the normality-based Prasad-Rao MSPE estimator
leads to underestimation, some times as large as about $8\%$ for
$m=60$ and about $10\%$ for $m=30.$ On the other hand, the
proposed MSPE estimator corrects for the underestimation in all
cases.  We note that for $m=30$, our proposed robust MSPE
estimator could lead to overestimation; but the overestimation
decreases dramatically as $m$ increases to 60. In contrast, the
underestimation for the Prasad-Rao estimation decrease very
slowly.  To highlight this point, consider the case where both  $\{e_i\}$
and $\{v_i\}$ follow the location shifted exponential distribution.  In this
case, as we increase $m$ from 30 to 60, underestimation for the
Prasad-Rao MSPE estimator decreases slowly from about $10\%$ to
$8\%$; in contrast, overestimation for the robust MSPE estimator
decreases dramatically from about $15\%$ to $4\%.$

Table 2 reports the percent relative root mean squared error
(RRMSE) of the MSPE estimators for the balanced case.  It shows
that the proposed MSPE estimator performs the best, in terms of
RRMSE, when the sampling errors $\{e_i\}$ are non-normal, while the
RRMSE in the normal  $\{e_i\}$ case are almost identical for the
proposed and the Prasad-Rao estimators.  The naive MSPE estimator
leads to relatively large RRMSE due to the large squared bias.

Tables 3 and 4 report the percent RB and RRMSE for the unbalanced
case and $m=60$ when $\psi$ is estimated by the Prasad-Rao simple
method-of-moments  and the Fay-Herriot method-of-moments,
respectively.  As in the balanced case, the naive MSPE estimator
underestimates the true MSPE in all situations, and the extent of the 
underestimation depends on the method of estimation of $\psi$ as
well as the distribution combination of the sampling errors ${\{ e_i\}}$ and
the random effects ${\{ v_i\}}$. It appears that for both balanced and
unbalanced cases, non-normality of ${\{e_i\}}$ causes more underestimation
than that of ${\{ v_i\}}$.

Table 3 shows that for the unbalanced case the RB for the
Prasad-Rao normality-based MSPE estimator is negligible whenever
the sampling errors $\{e_i\}$ are normally distributed regardless
of the distribution of ${\{v_i\}}$; this is again consistent with
previous theory (Lahiri and Rao, 1995).  Our theory in section 4
suggests that the normality-based Datta-Rao-Smith MSPE estimator,
unlike the Prasad-Rao MSPE estimator, is not second-order unbiased
even under the normality of ${\{ e_i\}}$.  It is, however, interesting to
note from Table 4 that there is virtually no difference between
the Datta-Rao-Smith estimator and the proposed robust estimator
(\ref{mspefh}) when ${\{ e_i\}}$ are normal. Note that the there are minor differences
between these two estimators in terms of RB even when both \{ $e_i$\} and
\{ $v_i$\} are normally distributed; but this difference can be
attributed to the fact that even in this case these two formulas
differ simply because we are using an estimator of the kurtosis of
the normal distribution in the robust MSPE formula (\ref{mspefh}).

The situation, however, is quite different when the sampling
errors $e_i$ are not normally distributed. The normality-based
Prasad-Rao and Datta-Rao-Smith MSPE estimators both tend to
underestimate, although in some non-normal situations both of them
perform well. It appears that the underestimation is more
prominent in the Prasad-Rao MSPE estimator than in the
Datta-Rao-Smith MSPE estimator. As in the balanced case, the
distribution combination of $\{ e_i \}$ and $\{ v_i\}$ is an important factor in
determining the extent of underestimation in both Prasad-Rao and
Datta-Rao-Smith MSPE estimators.  For example, when both $e_i$ and
$v_i$ follow the location shifted exponential distribution, there
is about an $8\%$ underestimation for the Datta-Rao-Smith MSPE
estimator for $G_1$; the corresponding underestimation for the Prasad-Rao
MSPE estimator being about $10\%$. We also note that there is
usually more underestimation for group $G_1$, the group with the smallest
ratio $\psi/D_i$ than for the other groups.

As $m$ increases, underestimation or overestmation for all the
methods decreases since they are all first-order unbiased.
However, underestimation for both the naive and the
normality-based Prasad-Rao or Datta-Rao-Smith MSPE estimator tend
to decrease at a slower rate than for our robust method - thus
verifying our theory. For example, consider group $G_1$ and the
case when $\{e_i\}$ and $\{v_i\}$ follow the location shifted exponential
distribution. When we increase $m$ from 60 to 100 for group $G_1$, underestimation
for the Prasad-Rao MSPE estimator decreases from about $-10\%$ to
$-7\%$. We note that for $m=60,$ our robust MSPE estimator suffers
from a very minor overestimation of about $2\%$. When we increase
$m$ to 100, overestimation drops to $0.25\%$.  For a small $m$, the
performance of the robust MSPE estimator depends on the estimator
of $\psi$ used. For $m=30$,  we have observed that the robust MSPE
estimator performs better when the Prasad-Rao method is used in
place of the Fay-Herriot method of estimating $\psi$. This is
possibly due to the fact that, unlike the Prasad-Rao method, we
need to estimate the kurtosis of $\{v_i\}$ when the Fay-Herriot method is
used and the estimation of kurtosis could be problematic for small
$m$.

Tables 3 and 4 also show that overall our proposed MSPE estimator
performs the best, in terms of RRMSE, when the sampling errors are
non-normal, while the RRMSE in the normal case are almost
identical for the proposed and the Prasad-Rao or DRS MSPE estimators. The
naive MSPE estimator leads to relatively large RRMSE due to the large
squared bias.

\begin{center}  {\large 6. D}ISCUSSION \end{center}

The additional assumption of known sampling kurtosis $\kappa_{ei}$  warrants further discussion.
We will first outline a method for estimating the kurtosis of the sampling distribution of the direct estimator $\bar{y}_{iw} = \sum_{k\in s_i} w_k y_k/\sum_{k\in s_i} w_k$ of the area mean $\bar{Y}_i$, 
where $s_i$ is the set of units in area $i$ that are selected in the sample
and $w_k$ is the basic  design weight of unit $k$, i.e., reciprocal of the first-order inclusion probability $\pi_k$ of unit $k$. The extension to calibration weighting should be possible through linearization but is not attempted here. To achieve this, we treat the sample design as Poisson sampling, which simplifies the evaluation of both the fourth moment and its estimator. In the following, we note that  published papers (e.g., \cite{morales19}) are using a sampling variance estimator, which is identical to that based on Poisson sampling, without realizing this. For smoothing purposes, Poisson sampling assumption is likely to perform well, especially when the sampling fraction within area is not large. Subsequently, we smooth the estimated fourth moment and the second moment to obtain a smoothed estimator of the kurtosis. This smoothed kurtosis estimator is then treated as the true kurtosis, consistent with the approach used in the normality-based Fay-Herriot model for smoothing sampling variances.

Using ratio approximation for the variance, \cite{sar92} obtained an approximate variance estimator of the direct domain estimator.  Denoting $\pi_{kl}$ the second-order inclusion probability of units $k$ and $l$, their approximate variance estimator of $\bar{y}_{iw}$ is given by:
$$v(\bar y_{iw})=\hat N_{iw}^{-2}\sum_{(k,l)\in s_i} \frac {\pi_{kl}-\pi_k\pi_l}{\pi_{kl}}\left (\frac {y_k-\bar y_{iw}}{\pi_k}  \right) \left (\frac {y_l-\bar y_{iw}}{\pi_l}  \right),$$
where $\hat N_{iw}=\sum_{k\in s_i}w_k$ is an estimator of the population size $N_i$ of area $i$. 
\cite{morales19} proposed the following simplification: 
\begin{equation}\label{morales}
  v(\bar{y}_{iw}) = \hat{N}_{iw}^{-2} \sum_{k\in s_i} w_k (w_k - 1) (y_k - \bar{y}_{iw})^2,  
\end{equation}
where $w_k = \pi_k^{-1}$. They treat this variance estimator as the true variance without any smoothing.  

We note that the assumption regarding the second-order inclusion probabilities, as considered by \cite{morales19}, is satisfied for Poisson sampling. We will first derive the formula (\ref{morales}) under Poisson sampling, by letting $\delta_{k} = 1$ if population unit $k$ in domain $i$ is in the sample and zero otherwise. Note that $\mathbb{E}(\delta_k) = \pi_k$, $\text{Var}(\delta_k) = \pi_k (1 - \pi_k)$ and the indicator variables $\delta_k$ are independent under Poisson sampling, which implies $\text{Cov}(\delta_k, \delta_l) = 0, \;k\ne l$. Under the ratio approximation, we apply the formula for estimating the variance of the estimated domain total $\hat{Z}_{iw} = \sum_{k\in s_i} w_k z_{k}$ of the variable $(y_k - \bar{Y}_i)/N_i = z_{k}$. The variance of $\hat{Z}_{iw}$ is given by
\[
V(\hat{Z}_{iw}) = \mathbb{E}\left[ \sum_{k\in U_i} (\delta_k - \pi_k) z_{k} \pi_k^{-1} \right]^2 = \sum_{k\in U_i} \mathbb{E}(\delta_k - \pi_k)^2 z_{k}^2 \pi_k^{-2} = \sum_{k\in U_i} V(\delta_k) z_{k}^2\pi_k^{-2},
\]
noting that the indicator variables $\delta_k$ are independent, where the summation is over $U_i$, the set of  all the population units in the domain $i$. Now using $V(\delta_k) = \pi_k (1 - \pi_k)$, it follows that
\[
V(\hat{Z}_{iw}) = \sum_{k\in U_i} (1 - \pi_k) \pi_k^{-1} {{z_{k}^2}}.
\]

Hence, an estimator of variance is given by
\[
v(\hat{Z}_{iw}) = \sum_{k\in s_i} (1 - \pi_k) \pi_k^{-2} {{z_{k}^2}} = \sum_{k\in s_i} (w_k - 1) w_k {{z_{k}^2}}.
\]
Now replacing {{$z_{k}$}} by $\hat{z}_{k} = (y_k - \bar{y}_{iw}) / \hat{N}_{iw}$, the above formula reduces to the formula used by \cite{morales19}.

We can use the above approach for variance estimation to obtain the fourth moment of $\bar{y}_{iw}$ and its estimator. We have
\begin{eqnarray}\label{fourth}
\mu_4 (\bar{y}_{iw}) &=& \mathbb{E}\left[ \sum_{k\in U_i} (\delta_k - \pi_k) {{z_{k}}} / \pi_k \right]^4 = \mathbb{E}\left[ \sum_{k\in U_i} (\delta_k - \pi_k)^2 {{z_{k}^2}}/ \pi_k^2 + 2 \sum_{k < l\in U_i} (\delta_k - \pi_k) (\delta_l - \pi_l) ({{z_{k}}} / \pi_k) ({{z_{l}}} / \pi_l) \right]^2\nonumber\\
&=& \sum_{k\in U_i} \mathbb{E}(\delta_k - \pi_k)^4 {{z_{k}^4}} / \pi_k^4 + 6 \sum_{k < l\in U_i} \mathbb{E}(\delta_k - \pi_k)^2 \mathbb{E}(\delta_l - \pi_l)^2 ({{z_{k}^2}} / \pi_k^2) ({{z_{l}^2}} / \pi_l^2), 
\end{eqnarray}
using independence of the indicator variables. We also have 
$$\mathbb{E}(\delta_k - \pi_k)^4 = \pi_k (1 - \pi_k) [\pi_k^3 + (1 - \pi_k)^3], \;\;\mathbb{E}(\delta_k - \pi_k)^2 = \pi_k (1 - \pi_k).$$

An estimator for the fourth moment is derived from equation (\ref{fourth}) by making the following modifications: replace $\sum_{k\in U_i}$ with $\sum_{k\in s_i}$, replace $\sum_{k < l\in U_i}$ with $\sum_{k < l \in s_i}$, multiply the term in the first sum by $\pi_k^{-1}$, multiply the term in the second sum by $\pi_k^{-1} \pi_l^{-1}$, and substitute {{$z_{k}$ with $\hat{z}_{k}$.}}  Once we have estimated the variance and  fourth central moment of $\bar{y}_{iw}$, we can obtain an estimator for the kurtosis of $\bar{y}_{iw}$.

To stabilize the sampling variance estimates, various smoothing techniques are available in the small area literature; see, for example, \cite{FH1979}, Chapter 6 of \cite{rao2015small}, \cite{HawalaLahiri2018}, \cite{lesageetal2022}, and \cite{You2023ApplicationOS}.  Essentially, variance smoothing is done by modeling the sampling variance estimate conditional on a set of relevant auxiliary variables. Often, a log linear model is considered. \cite{beaumont24} used random forests to model the estimated sampling variances as a function of area level covariates. This method makes it easier because no specification of the functional form is needed.  While the literature on variance smoothing can offer some insights into possible ways to smooth estimated sampling kurtosis, this is a new area of research that needs careful attention. 

Smoothed estimate of kurtosis of the sampling error could be helpful in choosing between normality-based Fay-Herriot model and the proposed semi-parametric Fay-Herriot model. However, in the normal case, the proposed MSPE estimator performs similarly to the Prasad-Rao MSPE estimator.


\begin{center}  {\large A}CKNOWLEDGEMENT \end{center}

S. Chen was supported in part by a Professional Development Award
from the RTI International. J.N.K. Rao was supported by a grant
from the Natural Sciences and Engineering Research Council of
Canada.  We thank Ms.Huilin Li for computational
support and the Editor and two anonymous referees for their helpful comments and constructive suggestions.

\setcounter{equation}{0}
\renewcommand{\theequation}{A.\arabic{equation}}

\begin{center}  {\bf \large{A}PPENDIX} \end{center}

The derivation of our MSPE estimator for $\hat\psi=\hat\psi_{PR}$
follows along the lines of the proof given in \cite{lahiri1995robust}.
Therefore, we focus on the proof for $\hat\psi=\hat\psi_{FH}.$

\noindent{\it Derivation of the asymptotic variance of
$\hat\psi$}:

Expanding $A(\hat\psi)$ around $\psi$ by Taylor series, we
have
\begin{eqnarray*}
\frac{1}{m-p}{\bf Y}'{\bf Q}(\psi){\bf Y}-1+\frac{1}{m-p}{\bf Y}'{\bf Q}^{(1)}({\bf \psi}){\bf Y}(\hat\psi- \psi)+O(\parallel \hat\psi-\psi\parallel^2)=0,
\end{eqnarray*}
where $A(\psi)$ is given by (2) and ${\bf Q}^{(j)}(\psi)$ is the $j$th
derivative of ${\bf Q}(\psi)$ with respect to $\psi~ (j=1,2,\cdots).$
Thus,
\begin{eqnarray*}
\hat\psi- \psi
=-\frac{\frac{1}{m-p}{\bf Y}'{\bf Q}(\psi){\bf Y}-1}{\frac{1}{m-p}{\bf Y}'{\bf Q}^{(1)}(\psi){\bf Y}}+O_p(m^{-1}),
\end{eqnarray*}
noting that $\hat\psi-\psi=O_p(m^{-\frac{1}{2}}).$

Define ${\bf v}=(v_1,\cdots,v_m)^{\prime},~{\bf e}=(e_1,\cdots,e_m)^{\prime},$
and ${\boldsymbol\varepsilon}={\bf v}+{\bf e}.$  Consider the transformation:
$\tilde{{\bf v}}={\bf \Sigma}^{-\frac{1}{2}}{\bf v},$
$\tilde{{\bf e}}={\bf \Sigma}^{-\frac{1}{2}}{\bf e},$
$\tilde{{{\boldsymbol\varepsilon}}}={\bf \Sigma}^{-\frac{1}{2}}{{\boldsymbol\varepsilon}}$, and
$\tilde{\bf X}={\bf \Sigma}^{-\frac{1}{2}}{\bf X}$.  Then, it is easy to see that
${\bf Y}'{\bf Q}(\psi){\bf Y}=\tilde{{{\boldsymbol\varepsilon}}}'\tilde{{\bf B}}\tilde{{{\boldsymbol\varepsilon}}},$
where
$\tilde{{\bf B}}={\bf I}-\tilde{{\bf X}}\{\tilde{{\bf X}}'\tilde{{\bf X}}\}^{-1}\tilde{{\bf X}}'.$
Since
$$\frac{1}{m-p}{\bf Y}'{\bf Q}^{(1)}(\psi){\bf Y}=-\frac{1}{m-p}tr({\bf \Sigma}^{-1})+O_p(m^{-\frac{1}{2}})$$
we obtain
\begin{eqnarray}
\hat\psi- \psi=\frac{1}{tr({\bf \Sigma}^{-1})}
\left[\tilde{{{\boldsymbol\varepsilon}}}'\tilde{{\bf B}}\tilde{{{\boldsymbol\varepsilon}}}-E(\tilde{{{\boldsymbol\varepsilon}}}'\tilde{{\bf B}}\tilde{{{\boldsymbol\varepsilon}}})\right]
+O_p(m^{-1}).
\end{eqnarray}
Now use (A.1) and apply part (d) of Lemma C.4 of Lahiri and Rao
(1995, p. 765) to get the asymptotic variance, $var(\hat\psi)$.

\noindent{\it Derivation of the second-order approximation to the
bias of $\hat\psi$}

Retaining the next term in the Taylor series expansion of
$A(\hat\psi)$, we obtain
\begin{eqnarray*}
\frac{{\bf Y}'{\bf Q}(\psi){\bf Y}}{m-p}-1+\frac{{\bf Y}'{\bf Q}^{(1)}(\psi){\bf Y}}{m-p}(\hat\psi- \psi)+\frac{1}{2}\frac{{\bf Y}'{\bf Q}^{(2)}(\psi){\bf Y}}{m-p}(\hat\psi-
 \psi)^{2}+O(\parallel \hat\psi-\psi\parallel^2)=0,
\end{eqnarray*}
which is equivalent to
\begin{eqnarray}
& &\frac{{\bf Y}'{\bf Q}(\psi){\bf Y}}{m-p}-1+  \frac{ {\bf Y}'{\bf Q}^{(1)}(\psi){\bf Y} - {\rm tr}
[{\bf \Sigma}(\psi){\bf Q}^{(1)}(\psi)]} {m-p}(\hat\psi- \psi)+ \frac{
{\rm tr} [{\bf \Sigma}(\psi){\bf Q}^{(1)}(\psi)]}{m-p}(\hat\psi-
\psi)\nonumber\\
 & &+ \frac{1}{2}\frac{{\bf Y}'{\bf Q}^{(2)}(\psi){\bf Y}- {\rm tr}[{\bf \Sigma}(\psi){\bf Q}^{(2)}(\psi)]}{m-p}(\hat\psi-
 \psi)^{2}+\frac{1}{2}\frac{{\rm tr}[{\bf \Sigma}(\psi){\bf Q}^{(2)}(\psi)]}{m-p}(\hat\psi-
 \psi)^{2}\nonumber\\
 & &= O(\parallel \hat\psi-\psi\parallel^2).
\end{eqnarray}
We now take expectation on both sides of (A.2). To this end, using
(A.1), Lemma C.4 of Lahiri and Rao (1995) and after considerable
algebra, we obtain
\begin{eqnarray}
E\left \{{\bf Y}'{\bf Q}^{(1)}(\psi){\bf Y} - {\rm tr}
[{\bf \Sigma}(\psi){\bf Q}^{(1)}(\psi)]\right \} (\hat\psi- \psi)&=&-2-
\frac{\psi^2\kappa_v{\rm tr}({\bf \Sigma}^{-3})+ {\rm tr}({\bf D}^2\Phi{\bf \Sigma}^{-3}) } {{\rm tr}({\bf \Sigma}^{-1})}  +o(1),\\
E\left \{{\bf Y}'{\bf Q}^{(2)}(\psi){\bf Y}- {\rm
tr}[{\bf \Sigma}(\psi){\bf Q}^{(2)}(\psi)]\right \}(\hat\psi-
 \psi)^{2}&=&o(1).
\end{eqnarray}

To obtain $b(\hat\psi;\psi,\kappa_v),$ the second-order
approximation to the bias of $\hat\psi$, we take the expectation in
(A.2), and then use (A.3), (A.4), the expression for $var(\hat\psi),$
and the following facts:
\begin{eqnarray*}
E[{\bf Y}'{\bf Q}(\psi){\bf Y}]&=&m-p\\
{\rm tr}[{\bf \Sigma}(\psi){\bf Q}^{(1)}(\psi)]&=&-{\rm tr}[{\bf Q}(\psi)]=-{\rm tr} [{\bf \Sigma}^{-1}(\psi)]+O(1) ;\\
{\rm tr}[{\bf Q}(\psi){\bf \Sigma}(\psi) {\bf Q}^{(1)}(\psi){\bf \Sigma}(\psi)]&=&-{\rm tr}[{\bf Q}(\psi)]=-{\rm tr} [{\bf \Sigma}^{-1}(\psi)]+O(1);\\
{\rm tr}[{\bf \Sigma} {\bf Q}^{(2)}(\psi)]&=&2{\rm tr}[{\bf Q}^{2}(\psi)]=2 {\rm tr}[{\bf \Sigma}^{-2}(\psi)]+O(1);\\
{\bf X}'{\bf Q}(\psi)&=&0.
\end{eqnarray*}

\noindent{\it Proof of Result 1}

Expanding  $\hat\theta_i(\hat\psi,{\bf Y})$ around $\psi$, we
obtain
\begin{eqnarray}
\hat\theta_i(\hat\psi,{\bf Y})-\hat\theta_i(\psi,{\bf Y})=(\hat\psi- \psi)\hat\theta_i^{(1)}(\psi,{\bf Y})+O_p(m^{-1}),
\end{eqnarray}
where $\hat\theta_i^{(1)}(\psi,{\bf Y})$ is the first derivative of
$\hat\theta_i(\psi,{\bf Y})$ with respect to $\psi$.  Write
\begin{eqnarray}
\hat\theta_i^{(1)}(\psi,{\bf Y})=\tilde{{\bf l}}'_i\tilde{{{\boldsymbol\varepsilon}}},
\end{eqnarray}
where $\tilde{{\bf l}}'_i=(\tilde{l}_{i1},\cdots,\tilde{l}_{im})$ with
$\tilde{l}_{ii}=\frac{D_i}{(\psi+D_i)^{3/2}}+O(m^{-1})$ and
$\tilde{l}_{ij}=O(m^{-1}),~j\ne i.$ Using (A.1) and (A.6) in
(A.5), we have
\begin{eqnarray}
E[\hat\theta_i(\hat \psi,{\bf Y})-\hat\theta_i(\psi, {\bf Y})]^2=\frac{1}{({\rm
tr}({\bf \Sigma}^{-1}))^2}E\left\{\left
(\tilde{{{\boldsymbol\varepsilon}}}'\tilde{{\bf B}}\tilde{{{\boldsymbol\varepsilon}}}-E[\tilde{{{\boldsymbol\varepsilon}}}'\tilde{{\bf B}}\tilde{{{\boldsymbol\varepsilon}}}]^2\right
) (\tilde{{\bf l}}'_i\tilde{{{\boldsymbol\varepsilon}}})^2\right\}+o(m^{-1}).
\end{eqnarray}
Application of Lemma C.4 of \cite{lahiri1995robust} to (A.7) and
algebra yields part (i) of Result 1.

To prove part (ii) of Result 1, write
\begin{eqnarray}
\hat\theta_i(\psi,{\bf Y})-\theta_i=\tilde{{\bf k}}'_i\tilde{{\bf v}}+\tilde{{\bf m}}'_i\tilde{{\bf e}},
\end{eqnarray}
where $\tilde{{\bf k}}'_i=(\tilde{k}_{i1},\cdots,\tilde{k}_{im}),\;
\tilde{{\bf m}}'_i=(\tilde{m}_{i1},\cdots,\tilde{m}_{im}),$ with
$\tilde{k}_{ii}=-\frac{D_i}{(\psi+D_i)^{1/2}}+O(m^{-1}),\;\tilde{m}_{ii}=\frac{\psi}{(\psi+D_i)^{1/2}}+O(m^{-1})$
and $\tilde{k}_{ij}=\tilde{m}_{ij}=O(m^{-1}),~j\ne i.$   Using
(A.1), (A.3),  (A.8) and the regularity conditions (r.1), (r.2)
and (r.3), we get
\begin{eqnarray}
&& E[\hat\theta_i(\hat \psi,{\bf Y})-\hat\theta_i(\psi,{\bf Y})][\hat\theta_i(\psi,{\bf Y})-\theta_i]\nonumber\\
&=&\frac{1}{tr({\bf \Sigma}^{-1})}E\{[\tilde{{\bf v}}'\tilde{{\bf B}}\tilde{{\bf v}}-E(\tilde{{\bf v}}'\tilde{{\bf B}}\tilde{{\bf v}}]
(\tilde{{\bf l}}'_i\tilde{{\bf v}})(\tilde{{\bf k}}'_i\tilde{{\bf v}})+[\tilde{{\bf e}}'\tilde{{\bf B}}\tilde{{\bf e}}-E(\tilde{{\bf e}}'\tilde{{\bf B}}\tilde{{\bf e}}]
(\tilde{{\bf l}}'_i\tilde{{\bf e}})(\tilde{{\bf m}}'_i\tilde{{\bf e}})\nonumber\\
&&+2(\tilde{{\bf v}}'\tilde{{\bf B}}\tilde{{\bf e}})(\tilde{{\bf l}}'_i\tilde{{\bf v}})(\tilde{{\bf m}}'_i\tilde{{\bf e}})
+2(\tilde{{\bf v}}'\tilde{{\bf B}}\tilde{{\bf e}})(\tilde{{\bf l}}'_i\tilde{{\bf e}})(\tilde{{\bf k}}'_i\tilde{{\bf v}})\}+o(m^{-1}).
\end{eqnarray}

Part (ii) of Result 1 follows from (A.9) after an application of
Lemma C.4 of Lahiri and Rao (1995) and algebra.

\bibliographystyle{chicago}
\bibliography{ref.bib}

\begin{thebibliography}{}

\bibitem[\protect\citeauthoryear{Beaumont, Bocci, Bosa, and Sombo}{Beaumont et~al.}{2024}]{beaumont24}
Beaumont, J.-F., C.~Bocci, K.~Bosa, and S.~Sombo (2024).
\newblock The use of random forests in small area estimation.
\newblock {\em Presented in the Seventh International Conference on Establishment Statistics (ICES VII) 2024\/}.

\bibitem[\protect\citeauthoryear{Chen and Lahiri}{Chen and Lahiri}{2008}]{cl08}
Chen, S. and P.~Lahiri (2008).
\newblock On mean squared prediction error estimation in small area estimation problems.
\newblock {\em Communications in Statistics -Theory and Methods\/}~{\em 37}, 1792--1798.

\bibitem[\protect\citeauthoryear{Das, Jiang, and Rao}{Das et~al.}{2004}]{DasJiangRao04}
Das, K., J.~Jiang, and J.~N.~K. Rao (2004).
\newblock Mean squared error of empirical predictor.
\newblock {\em The Annals of Statistics\/}~{\em 32\/}(2), 818--840.

\bibitem[\protect\citeauthoryear{Datta, Rao, and Smith}{Datta et~al.}{2005}]{Datta2005}
Datta, G., J.~Rao, and D.~Smith (2005).
\newblock On measuring the variability of small area estimators under a basic area level model.
\newblock {\em Biometrika\/}~{\em 92}, 183--196.

\bibitem[\protect\citeauthoryear{Datta and Lahiri}{Datta and Lahiri}{2000}]{datta2000}
Datta, G.~S. and P.~Lahiri (2000).
\newblock A unified measure of uncertainty of estimated best linear unbiased predictors in small area estimation problems.
\newblock {\em Statistica Sinica\/}~{\em 10}, 613--627.

\bibitem[\protect\citeauthoryear{Fay and Herriot}{Fay and Herriot}{1979}]{FH1979}
Fay, R. and R.~Herriot (1979).
\newblock Estimators of income for small area places: an application of james--stein procedures to census.
\newblock {\em Journal of the American Statistical Association\/}~{\em 74}, 269--277.

\bibitem[\protect\citeauthoryear{F\'uquene, Cristancho, Ospina, and Morales}{F\'uquene et~al.}{2019}]{morales19}
F\'uquene, J., C.~Cristancho, M.~Ospina, and D.~Morales (2019).
\newblock Prevalence of international migration: an alternative for small area estimation.
\newblock {\em arXiv\/}, 1905.00353.

\bibitem[\protect\citeauthoryear{Hawala and Lahiri}{Hawala and Lahiri}{2018}]{HawalaLahiri2018}
Hawala, S. and P.~Lahiri (2018).
\newblock Variance modeling for domains.
\newblock {\em Statistics and Applications (New Series)\/}~{\em 16\/}(1), 399--409.

\bibitem[\protect\citeauthoryear{Kackar and Harville}{Kackar and Harville}{1984}]{kac84}
Kackar, R. and D.~Harville (1984).
\newblock Approximations for standard errors of estimators of fixed and random effects in mixed linear models.
\newblock {\em Journal of the American Statistical Association\/}~{\em 79}, 853--862.

\bibitem[\protect\citeauthoryear{Lahiri and Rao}{Lahiri and Rao}{1995}]{lahiri1995robust}
Lahiri, P. and J.~Rao (1995).
\newblock Robust estimation of mean squared error of small area estimators.
\newblock {\em Journal of the American Statistical Association\/}~{\em 90\/}(430), 758--766.

\bibitem[\protect\citeauthoryear{Lesage, Beaumont, and Bocci}{Lesage et~al.}{2022}]{lesageetal2022}
Lesage, E., J.-F. Beaumont, and C.~Bocci (2022).
\newblock Two local diagnostics to evaluate the efficiency of the empirical best predictor under the fay-herriot model.
\newblock {\em Survey Methodology\/}~{\em 47\/}(2).

\bibitem[\protect\citeauthoryear{Otto and Bell}{Otto and Bell}{1995}]{otto1995sampling}
Otto, M.~C. and W.~R. Bell (1995).
\newblock Sampling error modelling of poverty and income statistics for states.
\newblock In {\em American Statistical Association, Proceedings of the Section on Government Statistics}, pp.\  160--165.

\bibitem[\protect\citeauthoryear{Pfeffermann and Nathan}{Pfeffermann and Nathan}{1981}]{pn81}
Pfeffermann, D. and G.~Nathan (1981).
\newblock Regression analysis of data from a cluster sample.
\newblock {\em Journal of the American Statistical Association\/}~{\em 76}, 681--689.

\bibitem[\protect\citeauthoryear{Prasad and Rao}{Prasad and Rao}{1990}]{prasad1990estimation}
Prasad, N. and J.~Rao (1990).
\newblock The estimation of the mean squared error of small-area estimators.
\newblock {\em Journal of the American statistical association\/}~{\em 85\/}(409), 163--171.

\bibitem[\protect\citeauthoryear{Rao and Molina}{Rao and Molina}{2015}]{rao2015small}
Rao, J. and I.~Molina (2015).
\newblock {\em Small area estimation}.
\newblock John Wiley \& Sons.

\bibitem[\protect\citeauthoryear{S\"arndal, Swensson, and Wretman}{S\"arndal et~al.}{1992}]{sar92}
S\"arndal, C.-E., B.~Swensson, and J.~Wretman (1992).
\newblock {\em Model Assisted Survey Sampling}.
\newblock New York, NY: Springer-Verlag.

\bibitem[\protect\citeauthoryear{Wolter}{Wolter}{1985}]{wol85}
Wolter, K.~M. (1985).
\newblock {\em Introduction to Variance Estimation}.
\newblock New York: Springer-Verlag Inc.

\bibitem[\protect\citeauthoryear{You and Hidiroglou}{You and Hidiroglou}{2023}]{You2023ApplicationOS}
You, Y. and M.~Hidiroglou (2023).
\newblock Application of sampling variance smoothing methods for small area proportion estimation.
\newblock {\em Journal of Official Statistics\/}~{\em 39}, 571 -- 590.

\end{thebibliography}

\newpage

\pagestyle{empty}

\noindent Table 1.  {\it Simulated values of percent relative bias
(RB) of mean squared prediction error estimators for the balanced
case: $\psi=D=1$ and $\psi$ is estimated by the Prasad-Rao (same as the Fay-Herriot)
method-of-moments.}

\vskip .2in

\begin{tabular}{llrrrrrr}
\multicolumn{8}{c}{Distribution of $v$}\\
& &\multicolumn{2}{c}{Normal} &\multicolumn{2}{c}{Double
Exponential}
&\multicolumn{2}{c}{Shifted Exponential}\\
Distribution of $e$& &$m=30$ &$m=60$ &$m=30$ &$m=60$ &$m=30$ &$m=60$\\
  ~\\
     &Naive &-12.10  &-6.64       & -6.66&-2.92            & -12.29 &-6.34      \\
    Normal &Prasad-Rao & 0.86       &-0.11   &0.53 &0.54 &1.61 & 0.54\\
     &Proposed &    0.86       &-0.11 &0.53 & 0.54 &1.61 &0.54\\
     ~\\

     &Naive &-17.90                     &-10.67                 &-9.60          & -4.96          &-17.59 & -10.82                      \\
     Double Exponential&Prasad-Rao & -5.35       & -4.22 & -2.59  &-1.52  & -4.21 &-4.16\\
     &Proposed &5.76     &1.10 & 1.82&  0.42&8.27&  1.53 \\
~\\
     &Naive &-22.13               &-14.17            &-11.91   & -6.67   &-22.65& -14.18          \\
  Shifted Exponential&Prasad-Rao & -9.86 &-7.90 & -4.85 &-3.25 &-9.69 &-7.60 \\
     &Proposed &12.4  &2.61 &4.33 & 0.66    &  15.05& 3.86
     \end{tabular}

\newpage
\vskip .9in

 \noindent Table 2.  {\it Simulated values of percent relative root mean
 squared
error (RRMSE) of the mean squared prediction error estimators for
the balanced case: $\psi=D=1$ and $\psi$ is estimated by the
Prasad-Rao (same as the Fay-Herriot) method-of-moments.}

\vskip .2in

\begin{tabular}{llrrrrrr}
\multicolumn{8}{c}{Distribution of $v$}\\
& &\multicolumn{2}{c}{Normal} &\multicolumn{2}{c}{Double
Exponential}
&\multicolumn{2}{c}{Shifted Exponential}\\
Distribution of $e$& &$m=30$ &$m=60$ &$m=30$ &$m=60$ &$m=30$ &$m=60$\\
~\\
     &Naive & 4.29         &   2.03                   &  2.48  & 1.05              &  5.99          &  3.20          \\
    Normal &Prasad-Rao & 2.60         &1.57   & 1.62 & 0.87  &  3.87 &  2.61\\
     &Proposed & 2.60           &1.57  &1.62  & 0.87  &  3.87 &  2.61 \\
~\\
     &Naive &  5.95                       &  2.95                &   3.04    &    1.34          & 7.59   &  4.20                \\
     Double Exponential&Prasad-Rao & 3.22      & 2.13  & 1.83   & 1.04   & 4.47   & 3.21  \\
     &Proposed & 1.77        & 1.57    & 1.14  &    0.86  &  2.61  &    2.39   \\
~\\
     &Naive &   7.58                    &   3.88            &  3.72               &    1.61        &  9.30           & 5.02                \\
  Shifted Exponential&Prasad-Rao & 4.04   &   2.75 &  2.18  & 1.21   &  5.25  &  3.73  \\
     &Proposed &  1.61  &   1.41   & 0.80    &    0.80    &    2.24   &1.99
     \end{tabular}

\newpage

\pagestyle{empty}

\landscape \noindent Table 3.  {\it Percent Relative Bias (
Relative Root Mean Squared Error) of the mean squared prediction
error estimators for the unbalanced case and $m=60$ when $\psi$ is
estimated by the Prasad-Rao simple method-of-moments}

\vskip .2in

\begin{tabular}{llrrrrrrrrr}
 & &\multicolumn{9}{c}{Dist. of $v$}\\
 & &\multicolumn{3}{c}{Normal} &\multicolumn{3}{c}{Double Exp.}
&\multicolumn{3}{c}{Shifted Exp.}\\
Dist. of $e$& Group &Naive &PR &Prop. &Naive &PR &Prop.&Naive &PR &Prop. \\
~\\
&G1 &-4.68 (3.49) & 0.19 (3.34) & 0.19 (3.34)&-3.11 (2.96) & 0.13 (2.83) &  0.13 (2.83)&-4.97 (6.00) &-0.03 (5.84) &-0.03 (5.84)\\
&G2 &-5.97 (0.82) &-0.33 (0.51) &-0.33 (0.51)&-2.43 (0.37) & 0.15 (0.27) &  0.15 (0.27)&-5.41 (1.43) & 0.84 (0.96) & 0.84 (0.96)\\
Normal&G3 &-5.78 (0.62) &-0.29 (0.34) &-0.29 (0.34)&-2.33 (0.25) &0.05 (0.18) &0.05 (0.18)&-6.54 (1.12)&-0.41 (0.62) &-0.41 (0.62)\\
&G4 &-5.48 (0.43) &-0.24 (0.20) &-0.24 (0.20)&-2.17 (0.15) &-0.06 (0.10) & -0.06 (0.10)&-4.71 (0.74) & 1.46 (0.37) & 1.46 (0.37) \\
&G5 &-3.97 (0.11) & 0.12 (0.03) & 0.12 (0.03)&-0.61 (0.03) & 0.77 (0.02) &  0.77 (0.02)&-4.66 (0.22) & 0.77 (0.05) & 0.77 (0.05) \\

~\\

&G1&-10.88 (5.13) &-6.31 (4.56) &-0.38 (4.03)&-5.12 (3.42) &-1.95 (3.15) &1.16 (2.88)&-10.61 (7.55) & -5.90 (7.01)  &0.24 (6.40)\\
Double&G2& -8.50 (1.22) &-2.78 (0.72) &1.68 (0.43)&-2.35 (0.42) &0.26 (0.31) &1.55 (0.26)&-10.35 (2.03) &-4.05 (1.20) &1.32 (0.63)\\
Exp.&G3& -9.03 (0.97) &-3.42 (0.49) &  0.79 (0.24)&-3.15 (0.32) &-0.74 (0.22) &0.37 (0.17)&-9.00 (1.53) &-2.54 (0.80) &2.86 (0.38)\\
&G4&  -6.80  (0.61) &-1.27 (0.27) & 2.75 (0.14)&-1.94  (0.19) & 0.21 (0.13) &1.14 (0.11)& -8.66 (1.10) &-2.16 (0.46) & 3.20 (0.20)\\
&G5& -4.44  (0.16) &0.12 (0.04) & 3.39 (0.12)&-1.69  (0.04) &-0.33 (0.03) &0.22 (0.02)& -7.14 (0.35) &-0.75 (0.07) &  4.88 (0.63) \\

~\\

&G1 &-13.28 (6.11) &-8.79 (5.37) &2.89 (4.32)&-7.67 (3.92) &-4.56 (3.47)  & 1.53 (2.79)&-14.44 (8.85) &-9.94 (8.04) &  1.77 (6.57)\\
Shifted&G2 &-10.33 (1.46) &-4.56 (0.84) &4.63 (0.34)&-3.77 (0.49) &-1.15 (0.35)  & 1.45 (0.22)&-11.05 (2.28) &-4.67 (1.36) &  6.41 (0.50)\\
Exp.&G3 &-10.78 (1.18)& -5.10 (0.60)&3.63 (0.18)& -4.20 (0.36) &-1.78 (0.23) &0.46 (0.13)&-11.66 (1.84) &-5.21 (0.94) &5.80 (0.30)\\
&G4 & -9.49 (0.80) &-3.93 (0.35) & 4.42 (0.15)&-4.68 (0.26) &-2.53 (0.16) &-0.67 (0.08)& -9.71 (1.27) &-3.05 (0.52) &  8.31 (0.46)\\
&G5 & -6.34 (0.21) &-1.56 (0.05) & 5.41 (0.41)&-2.91 (0.06) &-1.51 (0.03) &-0.43 (0.02)& -8.55 (0.42) &-1.78 (0.08) & 10.68 (2.63)\\

\end{tabular}

\newpage
\vskip .9in

\noindent Table 4.  {\it Percent Relative Bias (Root Mean Squared
Error) of the mean squared prediction error estimators for the
unbalanced case and $m=60$ when $\psi$ is estimated by the
Fay-Herriot method-of-moments}

\vskip .2in

\begin{tabular}{llrrrrrrrrr}
 & &\multicolumn{9}{c}{Dist. of $v$}\\
 & &\multicolumn{3}{c}{Normal} &\multicolumn{3}{c}{Double Exp.}
&\multicolumn{3}{c}{Shifted Exp.}\\
Dist. of $e$& Group &Naive &DRS &Prop. &Naive &DRS &Prop.&Naive &DRS &Prop. \\
~\\
&G1&-3.55 (2.48) &-0.32 (2.51) & 0.58 (2.52)&-3.14 (2.73) &-0.49 (2.67) & 0.67 (2.70)& -4.80 (4.98) &-1.64 (5.06) & 0.69 (5.36)\\
&G2&-3.92 (0.53) &-0.24 (0.43) &-0.05 (0.42)&-2.45 (0.34) &-0.37 (0.28) &-0.27 (0.27)&-3.82 (1.05) &-0.01 (0.92) & 0.49 (0.09)\\
Normal&G3&-4.26 (0.40)&-0.73 (0.30) &-0.64 (0.30)&-2.39 (0.24)&-0.48 (0.19) &-0.48 (0.19)&-3.83 (0.78) &-0.12 (0.65) &0.10 (0.63)\\
&G4&-3.24 (0.26) & 0.11 (0.19) & 0.06 (0.19)&-1.06 (0.13) & 0.65 (0.11) & 0.57 (0.10)&-2.86 (0.51) & 0.72 (0.41) & 0.62 (0.04)\\
&G5&-2.45 (0.06) &-0.01 (0.04) &-0.26 (0.04)&-1.71 (0.03) &-0.64 (0.02) &-0.82 (0.02)&-2.22 (0.13) & 0.55 (0.08) &-0.13 (0.09)\\

~\\
&G1&-6.88 (3.15) &-3.77 (3.04) &-0.44 (3.12)&-5.39 (3.06) &-2.79 (2.89) & 0.08 (2.81)&-9.36 (5.91) &-6.36 (5.81) &-1.79 (6.06)\\
Double &G2&-5.12 (0.65) &-1.48 (0.51) & 0.86 (0.45)&-2.34 (0.36) &-0.25 (0.30) &  0.70 (0.27)&-5.53 (1.28) &-1.78 (1.08) & 1.02 (0.95)\\
Exp.   &G3&-4.38 (0.47) &-0.85 (0.36) &1.24 (0.31)&-2.45 (0.25) &-0.54 (0.19) & 0.19 (0.17)&-6.11 (0.97) &-2.47 (0.77) &-0.07 (0.63)\\
&G4&-3.97 (0.31) &-0.64 (0.22) & 1.13 (0.18)&-2.60 (0.16) &-0.91 (0.12) &-0.40 (0.10)&-4.54 (0.63) &-0.99 (0.49) & 0.99 (0.39)\\
&G5&-3.63 (0.08) &-1.19 (0.04) &-0.25 (0.03)&-2.53 (0.04) &-1.47 (0.03) &-1.34 (0.02)&-3.28 (0.17) &-0.44 (0.10) & 0.53 (0.06)\\

~\\
&G1 &-9.03 (3.83) &-6.00 (3.64) &-0.44 (3.78)&-7.26 (3.40) &-4.72 (3.13) &-0.25 (2.89)&-10.98 (6.60) &-8.04 (6.44) &-1.40 (6.84)\\
Shifted&G2 &-5.44 (0.78) & -1.80 (0.62) & 2.69 (0.51)& -4.30 (0.44) &-2.25 (0.34) & -0.47 (0.25)& -8.46 (1.53) & -4.80 (1.24) & 0.16 (0.93)\\
Exp.&G3 &-6.89 (0.64) &-3.42 (0.46) & 0.61 (0.31)&-3.85 (0.31) &-1.96 (0.23) & -0.50 (0.17)& -7.64 (1.13) &-4.02 (0.88) & 0.54 (0.61)\\
&G4 &-6.37 (0.44) &-3.08 (0.29) & 0.49  (0.17)&-2.85 (0.17) &-1.15 (0.13) &-0.03 (0.10)& -6.44 (0.75) &-2.91 (0.56) & 1.19 (0.35)\\
&G5 &-3.78 (0.10) &-1.28 (0.06) & 1.03  (0.03)&-1.46 (0.04) &-0.38 (0.03) & 0.06 (0.02)& -4.69 (0.20) &-1.81 (0.11) & 1.06 (0.04)

\end{tabular}
\endlandscape

\end{document}